# Observation of nonlocal ferron-drag thermoelectricity

Takuma Itoh[1,9], Takamasa Hirai[1,2,*], Ping Tang[3,4], Hossein Sepehri-Amin[1], Ryo Iguchi[5], Yusuke Kozuka[5], Takao Shimizu[6], Soshi Akita[1], Shunsuke Mori[1,2], Gerrit E. W. Bauer[3,4,7,8] & Ken-ichi Uchida[1,2,*]

[1] Research Center for Magnetic and Spintronic Materials, National Institute for Materials Science, Tsukuba, Japan.

[2] Department of Advanced Materials Science, Graduate School of Frontier Sciences, The University of Tokyo, Kashiwa, Japan.

[3] Institute for Materials Research, Tohoku University, Sendai, Japan.

[4] WPI Advanced Institute for Materials Research, Tohoku University, Sendai, Japan.

[5] Research Center for Materials Nanoarchitectonics, National Institute for Materials Science, Tsukuba, Japan.

[6] Research Center for Electronic and Optical Materials, National Institute for Materials Science, Tsukuba, Japan.

[7] Center for Science and Innovation in Spintronics, Tohoku University, Sendai, Japan.

[8] Kavli Institute for Theoretical Sciences, University of the Chinese Academy of Sciences, Beijing, China.

[9] Present address: Global Research and Development Center for Business by Quantum-AI technology, National Institute of Advanced Industrial Science and Technology, Tsukuba, Japan.

[*] e-mail: HIRAI.Takamasa@nims.go.jp; UCHIDA.Kenichi@nims.go.jp

**The Peltier effect induces a heat current when a charge current passes through a conductor. Since a charge current is conserved at the junction between two different conductors, the difference between the heat flowing in both conductors for the same charge current leads to heating or cooling of the interface, providing an operating mechanism of solid-state heat pumps[1]. Here, we report observation of heat absorption and release signals even in a junction-free, homogeneous metal when**

**placed in proximity to a ferroelectric insulator. Our experiments using active thermographic imaging techniques confirm the prediction of the ferron-drag effect[2], i.e., the nonlocal excitation of ferrons, the collective excitation of the ferroelectric order[3], by conduction electrons in the adjacent metal. We reveal the electric-polarization-direction dependence of the temperature change signals and their unexpected increase with the metal thickness beyond the charge screening length, uncovering additional electron-phonon-ferron interactions in the metal/ferroelectric hybrid structure. The discovery of crosstalk between metals and ferroelectrics via remote ferrons could become both a nuisance and an opportunity for highly integrated circuits with ferroelectric barrier materials and revolutionize the design architecture of thermoelectric devices.**

Ferrons are quasiparticles representing collective excitation modes of electric polarization in ferroelectrics, i.e., the electric counterpart of magnons, collective excitations of the magnetic order. After the introduction of the concept in 2021 as carriers of energy and electric dipoles without carrying a net charge[4], the ferron theory has been developed in some detail[2-13]. Frequencies that can be tuned by external electric fields and relatively large group velocities render ferrons an attractive platform for unconventional thermoelectric transport[2,3,12]. The predicted thermoelectric conversion by ferrons in ferroelectrics[4-6] involves the interconversion between heat and electric dipole currents, which differs fundamentally from the conventional one in conductors caused by the asymmetric transport of conduction electrons and holes close to the Fermi energy[1], spin-caloritronic phenomena caused by the transport of electron spins and/or magnons[14,15], and dielectric Peltier effect induced by transient electric fields[16,17]. However, the experimental evidence of ferrons is still limited to a ferron contribution to the electric field dependence of the thermal conductivity of ferroelectrics[18], as well as the observation of coherent ferrons by THz spectroscopy[19-22] and diffuse ferron transport in microfabricated devices[23].

Ferron-induced thermoelectricity is not limited to the interior of ferroelectrics. In 2023, Tang et al. theoretically proposed a nonlocal ferron-drag effect that manifests itself in a layered structure of a ferroelectric insulator and an electrical conductor[2]. This phenomenon arises from momentum transfer between ferrons in the ferroelectric and conduction electrons in the conductor. Since ferrons and electrons live in different materials, conduction electrons drive ferron transport and the accompanying energy transfer in the adjacent ferroelectric "nonlocally" via the interlayer charge-dipole interaction (Fig. 1d). The nonlocal ferron-drag effect bridges ferron and electron transport phenomena, enabling access to the physics of ferron transport in insulators via well-established thermoelectric measurements of conductors.

Here we report the observation of the nonlocal ferron-drag thermoelectric effect. The conventional Peltier effect cannot be observed when a charge current is applied to a homogeneous conductor since the divergence of the induced heat current vanishes. The associated thermoelectric cooling or heating thus occurs only at heterojunctions comprising two dissimilar conductors (Fig. 1a,b). Contrary to this conventional wisdom, we observed charge-current-induced temperature changes when bonding an electrically insulating ferroelectric to part of a homogeneous metal film, despite the absence of any charge current in the ferroelectric (Fig. 1c). Through systematic experiments, we clarified that the origin of the temperature change signals stems from the nonlocal ferron-drag effect and demonstrated the existence of a "phonon-mediated" electron-ferron interaction that transcends the framework of conventional ferron transport theories. Our findings open new frontiers in the physics of ferroelectric devices and provide an unconventional strategy for thermal management based on energy transport in electrically insulating materials.

**Sample system and measurement procedures**

We demonstrate the nonlocal ferron-drag effect using a Pt film as the conductor and a $LiNbO_3$ substrate as the ferroelectric insulator. $LiNbO_3$ is a prominent ferroelectric insulator with high Curie temperature (~1200 °C) and large spontaneous electric polarization along the polar *c*-axis of the crystal structure. We first deposited a thin-film strip of dielectric insulator $SiO_2$ onto a single-crystal $LiNbO_3$ substrate, followed by depositing a Pt strip perpendicular to it (Fig. 2a). As depicted in Extended Data Fig. 1, multiple Pt and $SiO_2$ strips with different thicknesses on the $LiNbO_3$ substrate allows us to measure the Pt- and $SiO_2$-thickness dependences of thermoelectric responses while minimizing variations between samples (see Methods for details). At the intersections, the $SiO_2$ film acts as a spacer barrier, whereas in other areas, Pt is in direct contact with $LiNbO_3$. We do not observe any atomic diffusion at the Pt/$LiNbO_3$ interface (Extended Data Fig. 2), presumably because the sample preparation does not involve any heating. The main experiments below were conducted using *Z*-cut $LiNbO_3$ substrates with electric polarization **P** uniformly aligned perpendicular to the Pt/$LiNbO_3$ interface (*z* direction).

We studied the thermoelectric response of the Pt/$LiNbO_3$ samples by active infrared emission microscopy, also called lock-in thermography (LIT)[24-28]. The LIT method accesses highly resolved spatiotemporal temperature modulations by measuring the infrared radiation emitted from the surface of the samples using an infrared camera. We focused on the response to a square-wave-modulated alternating charge current with amplitude $J_c^{sq}$ and frequency $f$ (= 25 Hz) applied to the Pt layer along the $x$ direction, where the charge current has no net offset from zero (Fig. 2a,c). By extracting the first harmonic response of the detected thermal images, we can separate the thermoelectric signal ($\propto J_c^{sq}$) from the Joule heating signal ($\propto (J_c^{sq})^2$) because the latter does not depend on the current direction and therefore does not have oscillating components (Fig. 2b). We present the results in terms of spatial images of the lock-in amplitude $A$ and phase $\phi$ with $A \geq 0$ and $0° \leq \phi < 360°$ (Fig. 2a). The former represents the magnitude and the latter the sign of the temperature modulation that does not show significant time delay in thin-film-based devices[27,28]. At the relatively high frequency $f$ used here, the temperature broadening due to thermal diffusion is suppressed, enabling accurate determination of positions of the heat sources and sinks[26]. The calibration detailed in ref. 26 allows us to convert the observed infrared radiation intensity into temperature distributions. Dusing LIT measurements, the steady-state temperature of the samples was fixed near room temperature using a Peltier module at atmospheric pressure.

**Emergence of thermoelectric signals caused by proximate ferroelectric**

Figure 2d,e shows the $A$ and $\phi$ images for a Pt/*Z*-cut $LiNbO_3$ sample with a Pt thickness $t_{Pt}$ of 12 nm and a $SiO_2$ thickness of 7 nm at $J_c^{sq}$ = 8.0 mA. Although the charge current flows only in the Pt layer with a high electron density and electronic structure that is not affected by the proximity of a ferroelectric insulator, distinct temperature changes were observed at the edges of the Pt/$SiO_2$ overlap region. Figure 2f shows a line profile of $A$ along the charge current direction: the temperature changes peak precisely at the boundaries between the area where Pt is in direct contact with $LiNbO_3$ and that where it is not. A phase shift of 180° between the left boundary (position L) and the right boundary (position R) implies that a charge current along the $-x$ direction heats the L edge and cools the R edge (Fig. 2e,g). We confirmed that **P** reversal has no effect on the magnitude and sign of the temperature changes. Figure 2h,i shows the $J_c^{sq}$ dependence of the temperature changes at the boundaries: $A$ increases proportional to $J_c^{sq}$, while $\phi$ is constant. This behavior confirms the thermoelectric origin of the temperature change signals that requires a different ratio of heat and charge currents between the two areas. Since the charge current is continuous across the entire Pt layer, the heat flux on the left side of the L boundary is larger than that to the right (left of the R boundary smaller than that to its right). The observed thermoelectric signals depend only slightly on the thickness of the $SiO_2$ layer, indicating that the direct contact between Pt and $LiNbO_3$ is the decisive factor (Extended Data Fig. 3). The sign of the thermoelectric signals in the Pt/*Z*-cut $LiNbO_3$ sample is consistent with the nonlocal ferron-drag effect to be discussed below.

We carried out several control measurements to exclude other explanations. The importance of the ferroelectric order was tested by a sample in which the $LiNbO_3$ substrate was replaced with a single-crystal *Z*-cut quartz with similar crystal structure to *Z*-cut $LiNbO_3$ but without spontaneous electric polarization. The fabrication procedures and measurement conditions were the same as those for the Pt/*Z*-cut $LiNbO_3$ sample. As shown in the $A$ and $\phi$ images in Fig. 3a,b, thermoelectric signals in the Pt/*Z*-cut quartz sample are finite but much smaller than those in the Pt/*Z*-cut $LiNbO_3$ sample with the same $t_{Pt}$. The maximum amplitude $A_p$ obtained for the Pt/*Z*-cut quartz sample at each $J_c^{sq}$ was only ~13% of that for the Pt/*Z*-cut $LiNbO_3$ sample (Fig. 3c), a difference that cannot be explained by bulk thermophysical properties of the substrates, such as the difference in out-of-plane thermal conductivities of *Z*-cut quartz (12.7 $Wm^{-1}K^{-1}$) and *Z*-cut $LiNbO_3$ (4.4 $Wm^{-1}K^{-1}$) (Extended Data Fig. 4). The computed temperature changes induced by the same heat sources are similar for both substrates because the thermal conductivity difference is compensated for by the different volumetric specific heats (see Extended Data Fig. 5 and Methods). The nonvanishing signals in the Pt/*Z*-cut quartz sample may be evidence of the conventional phonon drag, i.e., the momentum transfer from conduction electrons in the Pt film to remote polar phonons[29-33] in the adjacent quartz, which is unpolarized in the absence of strain and voltage. The smallness of the signals can be explained by the small group velocities of the polar phonons. Based on Fig. 3, we conclude that ferroelectricity enhances the thermoelectric effects by an order of magnitude under otherwise very similar conditions.

**Long-range nature and polarization dependence of thermoelectric signals**

Next, we focus on the dependence of the observed thermoelectric signals in the Pt/*Z*-cut $LiNbO_3$ samples on the Pt thickness for a $SiO_2$-spacer thickness of 7 nm. It is natural to expect that the observed thermoelectric signal is an interfacial effect, since electric polarization is efficiently screened by conduction electrons in Pt and $A_\mathrm{p}/j_\mathrm{c}^\mathrm{sq}$, the peak amplitude normalized by the applied charge current density $j_\mathrm{c}^\mathrm{sq}$, should not depend on $t_\mathrm{Pt}$. However, as shown with the purple circle data points in Fig. 4a, the magnitude of $A_\mathrm{p}/j_\mathrm{c}^\mathrm{sq}$ continues to increase even when $t_\mathrm{Pt}$ is on the order of tens of nanometers, a length scale that exceed the Thomas-Fermi screening length in Pt (~0.05 nm)[34] by three orders of magnitude.

The thermoelectric signals were found to depend on the crystal orientation or **P** direction of the ferroelectric. Figure 4 summarizes the results of LIT measurements on samples in which the *Z*-cut $LiNbO_3$ substrate was replaced by an *X*-cut one with in-plane electric polarization along the Pt strips. Although the thermoelectric signals did not change when **P** was rotated 180° in the Pt/*Z*-cut $LiNbO_3$ samples, the signal intensities in the *Z*-cut and *X*-cut samples were clearly different. The significantly smaller amplitudes in the Pt/*X*-cut $LiNbO_3$ samples for $t_\mathrm{Pt} < 70$ nm cannot be explained by the small difference in out-of-plane thermal conductivities of *Z*-cut $LiNbO_3$ (4.4 $\mathrm{Wm^{-1}K^{-1}}$) and *X*-cut $LiNbO_3$ (3.8 $\mathrm{Wm^{-1}K^{-1}}$) (Extended Data Figs. 4 and 5), nor by the difference of the ferroelectric stray fields on metals with a high electron density. Since the electrical resistivity of the Pt films on *Z*-cut $LiNbO_3$ is almost the same as that on *X*-cut $LiNbO_3$ (Extended Data Fig. 6), the observed dependence on the electric polarization is not due to the difference in the Pt film quality.

**Phonon-mediated ferron drag**

The original theory of the nonlocal ferron-drag effect[2] focused on two-dimensional van der Waals ferroelectric-semiconductor bilayers (e.g., $CuInP_2S_6$/graphene bilayers) must be modified to address the long-range nature of the thermoelectric signals in the Pt/$LiNbO_3$ samples. In atomic monolayers on ferroelectrics, ferrons can directly couple to conduction electrons by the interlayer charge-dipole interaction due to the reduced screening. In contrast, in our Pt/$LiNbO_3$ samples, the stray fields of the electric polarization emitted by $LiNbO_3$ are screened in Pt on a subatomic length scale[34], as discussed above.

The observed long-range nature of the thermoelectric signals leads us to consider a “phonon-mediated” electron-ferron interaction that is not as easily screened by conduction electrons. This idea can be formulated by the linearized Boltzmann equation in the relaxation time approximation with the electron-ferron interaction that acquires long-range nature via the coupling with an acoustic phonon bath (Methods). As illustrated in Fig. 1d, the conduction electrons in the metal emit (acoustic) phonons by the electron-phonon coupling that subsequently interact strongly with the ferrons in the adjacent ferroelectric via piezoelectric (or electrostrictive) couplings. Applying a charge current $J_{\mathrm{c}}$ therefore drives a ferron number and heat current $J_{\mathrm{q}}$ leading to an additional ferron-drag contribution, i.e.,

$$J_{\mathrm{q}} = (\Pi_{\mathrm{e}} + \Pi_{\mathrm{fd}})J_{\mathrm{c}} \quad (1)$$

where $\Pi_{\mathrm{e}}$ is the electronic Peltier coefficient of the Pt layer incorporating the conventional phonon-drag correction, while $\Pi_{\mathrm{fd}}$ is the nonlocal ferron-drag coefficient. The charge-current-induced heating and cooling signals are then caused by the divergence of $J_{\mathrm{q}}$ at the boundaries between the Pt/$LiNbO_3$ and Pt/$SiO_2$/$LiNbO_3$ regions. Since $\Pi_{\mathrm{fd}}$ is reduced by the $SiO_2$ spacer, we expect heating when electrons flow from Pt/$LiNbO_3$ to Pt/$SiO_2$/$LiNbO_3$ and cooling at the other edge, which agrees with observations. We have no indications of strong scattering at a Pt/$SiO_2$ interface that in principle could affect $\Pi_{\mathrm{e}}$ in very thin Pt films. The conventional phonon-drag contribution that might modify $\Pi_{\mathrm{e}}$ differently under the Pt/$LiNbO_3$ and Pt/$SiO_2$/$LiNbO_3$ regions can be ruled out because it should also operate in the Pt/quartz samples with much smaller thermoelectric signals (Fig. 3).

The importance of the electron-phonon interaction in the metallic layer in the phonon-mediated nonlocal ferron-drag effect was confirmed by control LIT measurements on a *Z*-cut $LiNbO_3$ sample in which the Pt strip was replaced by a Au strip. Since the electron-phonon interaction in Au is much smaller than that in Pt[35-39], the nonlocal ferron-drag signals should then be significantly suppressed. Indeed, Fig. 5 shows that for equal charge current densities in the Au strip, we could not resolve the amplitude of temperature changes near the edges of the $SiO_2$ strip (Fig. 5a,c) with random phases (Fig. 5b,d). Hence, the phonon-mediated nonlocal ferron-drag mechanism can qualitatively explain a sizable thermoelectric conversion in Pt/ferroelectric heterostructures despite the very short screening length of Pt.

The microscopic origin of the polarization-direction dependence of the thermoelectric signals in Fig.

4a remains unclear, but it could be due to anisotropic ferron group velocities in the *Z*-cut and *X*-cut $LiNbO_3$ substrates, as well as their anisotropic piezoelectric constants[40-42] (Methods). According to Fig. 4a, the polarization-direction dependence is largest for 40 < $t_{Pt}$ < 70 nm at which the signal of the Pt/*Z*-cut $LiNbO_3$ samples appears to saturate at a higher level than the *X*-cut samples, while for $t_{Pt}$ > 70 nm both signals merge and increase monotonically at the same rate. The "switchability" could be an important functionality in thermoelectric devices that deserves to be understood in follow-up theoretical and experimental studies.

## Conclusion

We report the thermoelectric signals induced by the contact between a metal and ferroelectric and interpret the observed phenomenon in terms of a ferron-drag effect that reaches far beyond the charge screening length. The experiments sprouted the idea of a phonon-mediated electron-ferron interaction, which we support by a phenomenological model explaining most observations. A phonon-mediated nonlocal ferron-drag effect enables the conversion between charge and ferron currents that is not constrained by charge screening. It should exist for any ferroelectric that is bonded to a conductor and therefore could be much larger for optimized material combinations. It may affect transport properties in semiconductor devices and integrated circuits with ferroelectric barrier materials[43-45]. Our findings are the first step toward establishing novel thermal management technologies utilizing insulators, since further systematic material screening and elucidation of the length scales involved in the electron-phonon-ferron interactions will improve the performance.

## References

1. Goldsmid, H. J. *Introduction to Thermoelectricity* (Springer, 2010).
2. Tang, P., Uchida, K. & Bauer, G. E. W. Nonlocal drag thermoelectricity generated by ferroelectric van der Waals heterostructures. *Phys. Rev. B* **107**, L121406 (2023).
3. Tang, P., Iguchi, R., Uchida, K. & Bauer, G. E. W. Excitations of the ferroelectric order. *Phys. Rev. B* **106**, L081105 (2022).

[4] Bauer, G. E. W., Iguchi, R. & Uchida, K. Theory of transport in ferroelectric capacitors. *Phys. Rev. Lett.* **126**, 187603 (2021).

[5] Bauer, G. E. W., Tang, P., Iguchi, R. & Uchida, K. Magnonics vs. ferronics. *J. Magn. Magn. Mater.* **541**, 168468 (2022).

[6] Tang, P., Iguchi, R., Uchida, K. & Bauer, G. E. W. Thermoelectric polarization transport in ferroelectric ballistic point contacts. *Phys. Rev. Lett.* **128**, 047601 (2022).

[7] Shen, K. Magnon-ferron coupling mediated by dynamical Dzyaloshinskii-Moriya interaction in a two-dimensional multiferroic model. *Phys. Rev. B* **106**, 104411 (2022).

[8] Adachi, H., Ikeda, N. & Saitoh, E. Ginzburg-Landau action and polarization current in an excitonic insulator model of electronic ferroelectricity. *Phys. Rev. B* **107**, 155142 (2023).

[9] Zhou, X.-H. et al. Surface ferron excitations in ferroelectrics and their directional routing. *Chinese Phys. Lett.* **40**, 087103 (2023).

[10] Shen, K. Electrical and magnetic control of spin-lattice configuration and magnon-ferron hybridization in a two-dimensional multiferroic model. *Phys. Rev. B* **108**, 094413 (2023).

[11] Bauer, G. E. W. et al. Polarization transport in ferroelectrics. *Phys. Rev. Appl.* **20**, 050501 (2023).

[12] Tang, P. & Bauer, G. E. W. Electric analog of magnons in order-disorder ferroelectrics. *Phys. Rev. B* **109**, L060301 (2024).

[13] Rodríguez-Suárez, R. L. et al. Surface and volume modes of polarization waves in ferroelectric films. *Phys. Rev. B* **109**, 134307 (2024).

[14] Uchida, K. & Hirai, T. Spin caloritronics: History and future prospects of experiments, *J. Magn. Magn. Mater.* **652**, 174195 (2026).

[15] Bauer, G. E.W., Saitoh, E. & van Wees, B. J. Recollections on spin caloritronics. *J. Magn. Magn. Mater.* **653**, 174115 (2026).

[16] Marvan, M. The electric polarization induced by temperature gradient and associated thermoelectric effects. *Czech J. Phys.* **19**, 1240-1245 (1969).

[17] Iguchi, R. et al. Thermoelectric response of a ferroelectric insulator. arXiv:2606.25767

[18] Wooten, B. L. et al. Electric field-dependent phonon spectrum and heat conduction in ferroelectrics. *Sci. Adv.* **9**, eadd7194 (2023).

[19] Choe, J. et al. Observation of coherent ferron emission and propagation. *Nat. Mater.* (2026) doi: 10.1038/s41563-026-02597-4

[20] Jana, S. et al. Ferron-driven photoferroic hysteresis in van der Waals $CuInP_2S_6$. *Nat. Commun.* **17**, 7100 (2026).

[21] Zhang, B. et al. Electric-field control of giant ferronics. arXiv:2509.06057

[22] Subedi, S. et al. Electrically switchable ferron upconversion in a van der Waals ferroelectric. arXiv:2603.19394

[23] Shen, K. et al. Observation of ferron transport in ferroelectrics. arXiv:2505.24419

[24] Breitenstein, O., Warta, W. & Langenkamp, M. *Lock-in Thermography: Basics and Use for Evaluating Electronic Devices and Materials.* (Springer, Berlin, Heidelberg, 2010).

[25] Daimon, S., Iguchi, R., Hioki, T., Saitoh, E. & Uchida, K. Thermal imaging of spin Peltier effect. *Nat. Commun.* **7**, 13754 (2016).

[26] Uchida, K., Daimon, S., Iguchi, R. & Saitoh, E. Observation of anisotropic magneto-Peltier effect in nickel. *Nature* **558**, 95-99 (2018).

[27] Seki, T., Iguchi, R., Takanashi, K. & Uchida, K. Visualization of anomalous Ettingshausen effect in a ferromagnetic film: Direct evidence of different symmetry from spin Peltier effect. *Appl. Phys. Lett.* **112**, 152403 (2018).

[28] Das, R., Iguchi, R. & Uchida, K. Systematic investigation of anisotropic magneto-Peltier effect and anomalous Ettingshausen effect in Ni thin films. *Phys. Rev. Appl.* **11**, 034022 (2019).

[29] Fischetti, M. V., Neumayer, D. A. & Cartier, E. A. Effective electron mobility in Si inversion layers in metal-oxide-semiconductor systems with a high-$\kappa$ insulator: The role of remote phonon scattering. *J. Appl. Phys.* **90**, 4587-4608 (2001).

[30] Laikhtman, B. & Solomon, P. M. Remote phonon scattering in field-effect transistors with a high $\kappa$ insulating layer. *J. Appl. Phys.* **103**, 014501 (2008).

[31] Fratini, S. & Guinea, F. Substrate-limited electron dynamics in graphene. *Phys. Rev. B* **77**, 195415 (2008).

[32] Koh, Y. K. et al. Role of remote interfacial phonon (RIP) scattering in heat transport across graphene/$SiO_2$ interfaces. *Nano Lett.* **16**, 6014-6020 (2016).

[33] You, Y. G. et al. Role of remote interfacial phonons in the resistivity of graphene. *Appl. Phys. Lett.* **115**, 043104 (2019).

[34] Scalfi, L. & Rotenberg, B. Microscopic origin of the effect of substrate metallicity on interfacial free energies. *Proc. Natl Acad. Sci. USA* **118**, e2108769118 (2021).

[35] Bardeen, J. & Pines, D. Electron-phonon interaction in metals. *Phys. Rev.* **99**, 1140-1150 (1955).

[36] Allen, P. B. Electron-phonon effects in the infrared properties of metals. *Phys. Rev. B* **3**, 305-320 (1971).

[37] Karvonen, J. T., Taskinen, L. J. & Maasilta, I. J. Electron-phonon interaction in thin copper and gold films. *Phys. Stat. Sol. (c)* **1**, 2799-2802 (2004).

[38] Xu, M., Yang, J.-Y., Zhang, S. & Liu, L. Role of electron-phonon coupling in finite-temperature dielectric functions of Au, Ag, and Cu. *Phys. Rev. B* **96**, 115154 (2017).

[39] Smirnov, N. A. Copper, gold, and platinum under femtosecond irradiation: Results of first-principles calculations. *Phys. Rev. B* **101**, 094103 (2020).

[40] Yamada, T., Niizeki, N. & Toyoda, H. Piezoelectric and elastic properties of lithium niobate single crystals. *Jpn. J. Appl. Phys.* **6**, 151-155 (1967).

[41] Smith, R. T. & Welsh, F. S. Temperature dependence of the elastic, piezoelectric, and dielectric constants of lithium tantalate and lithium niobate. *J. Appl. Phys.* **42**, 2219-2230 (1971).

[42] Weis, R. S. & Gaylord, T. K. Lithium niobate: Summary of physical properties and crystal structure. *Appl. Phys. A* **37**, 191-203 (1985).

[43] Wilk, G. D., Wallace, R. M. & Anthony, J. M. High-$\kappa$ gate dielectrics: Current status and materials properties considerations. *J. Appl. Phys.* **89**, 5243-5275 (2001).

[44] Mistry, K. et al. A 45nm logic technology with high-k+metal gate transistors, strained silicon, 9 Cu interconnect layers, 193nm dry patterning, and 100% Pb-free packaging. 2007 IEEE International Electron Devices Meeting, Washington, DC, USA, 247-250 (2007).

[45] Böscke, T. S., Müller, J., Bräuhaus, D., Schröder, U. & Böttger, U. Ferroelectricity in hafnium oxide thin films. *Appl. Phys. Lett.* **99**, 102903 (2011).

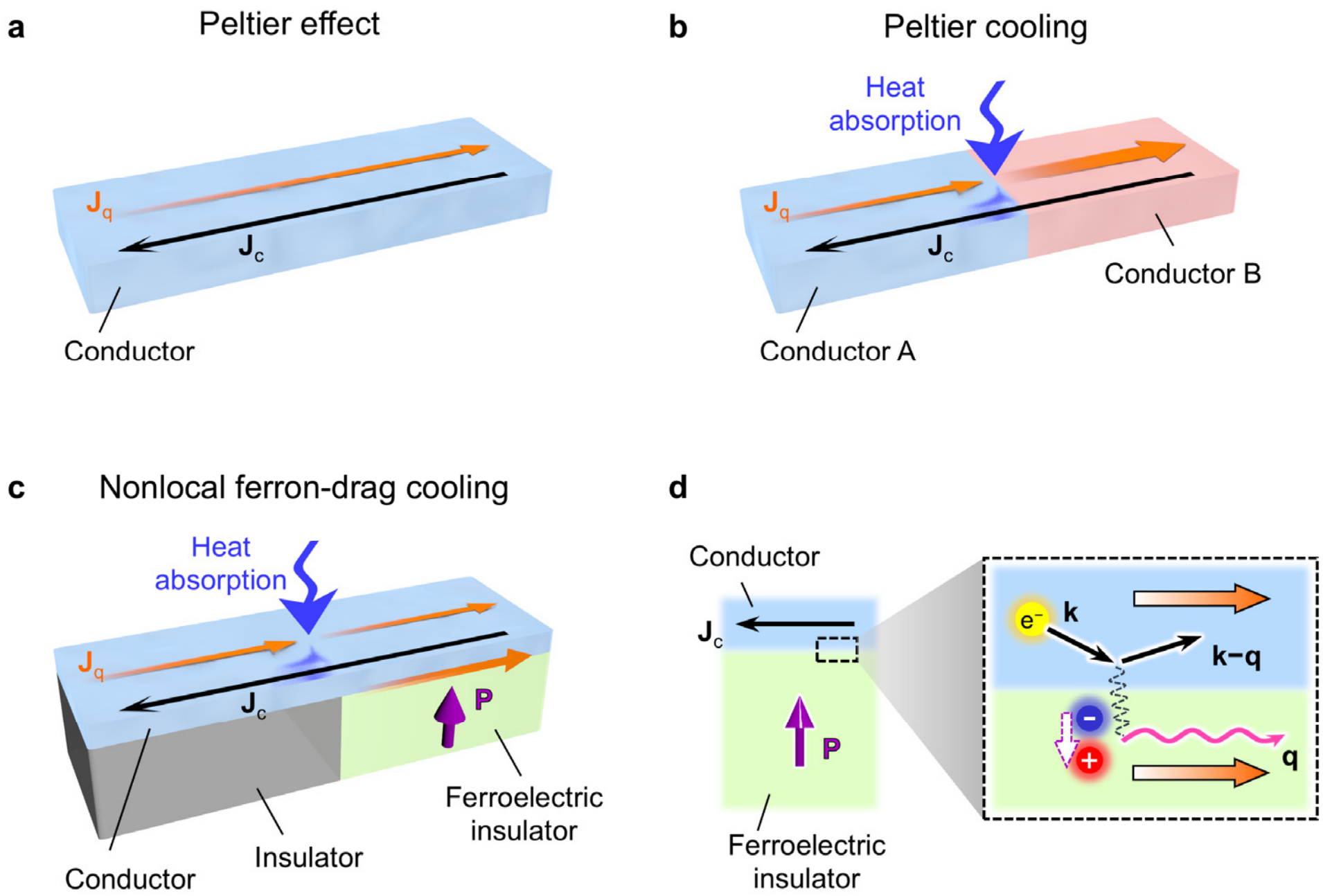

**Fig. 1 | Peltier cooling and nonlocal ferron-drag cooling. a,** Schematic of the Peltier effect in a single, homogeneous conductor. When a charge current $\mathbf{J}_c$ is applied to the conductor, a heat current $\mathbf{J}_q$ is generated along the $\mathbf{J}_c$ direction, where the ratio between the applied charge current and generated heat current is the Peltier coefficient. **b,** Schematic of Peltier cooling. When $\mathbf{J}_c$ flows through a junction comprising two conductors A and B with different Peltier coefficients, the conservation of energy causes heat absorption or release at the interface proportional to the discontinuity of the heat current. **c,** Schematic of nonlocal ferron-drag cooling. When a ferroelectric insulator is bonded to only a part of a conductor, the elementary excitations of the ferroelectric order are dragged by momentum transfer from the electrons in the conductor that generates a heat flow in the ferroelectric. This leads to excess heat absorption or release at the edges of the ferroelectric. **P** denotes the spontaneous electric polarization of the ferroelectric. **d,** Schematic of the electron-ferron interaction near the conductor/ferroelectric interface. Electrons in the conductor (black arrows) excite a ferron with a wave vector **q** (pink wavy line) by scattering from a wave vector **k** to **k**−**q**. The black wavy line represents the long-range interaction vertex mediated by virtual phonons.

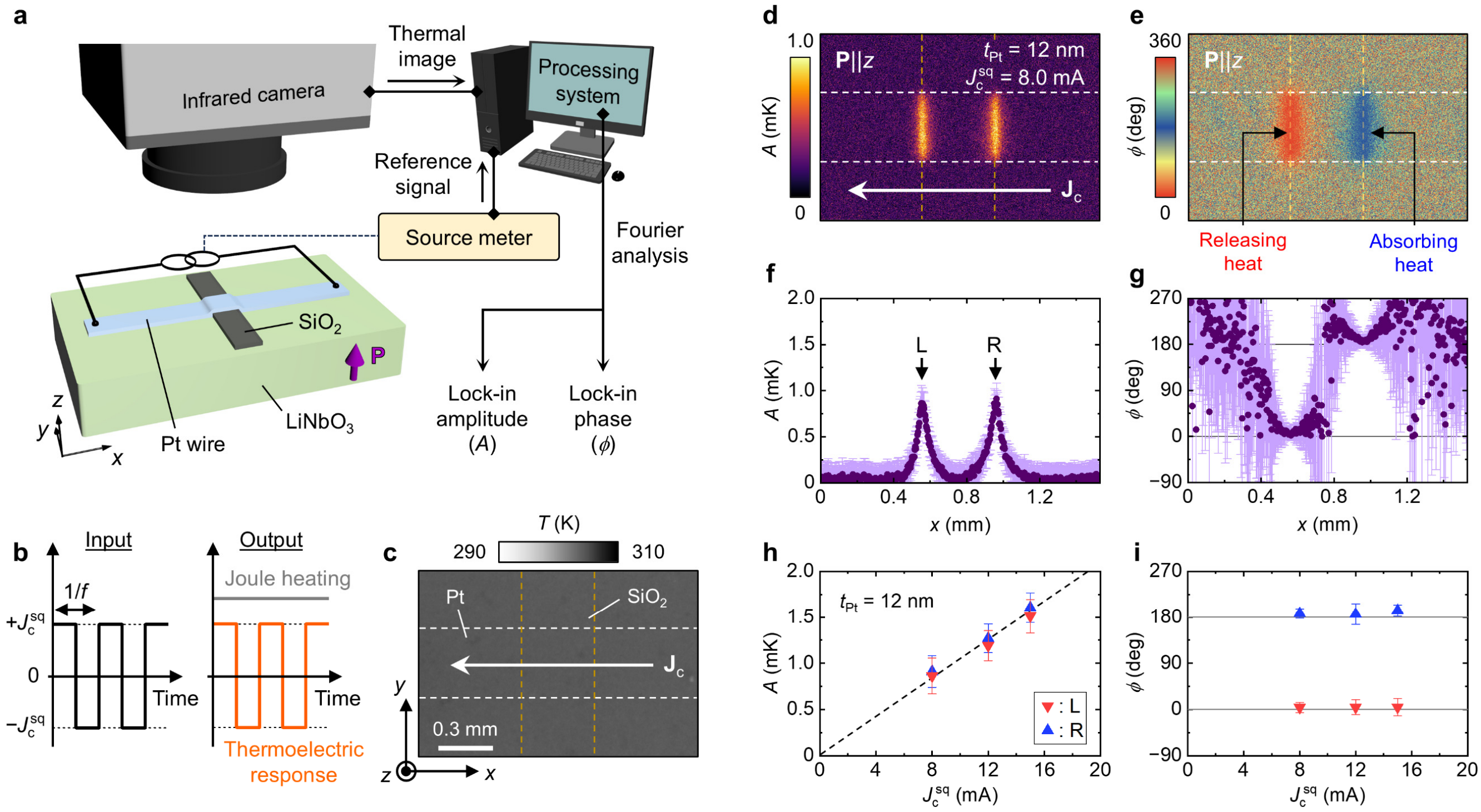

**Fig. 2 | Experimental set-up and thermoelectric response in Pt/$LiNbO_3$. a,** Schematic of the lock-in thermography (LIT) setup. The Pt and $SiO_2$ strips on the $LiNbO_3$ substrate are along the *x* and *y* directions, respectively. **b,** Square-wave-modulated input charge current and output temperature changes due to thermoelectric effects and Joule heating in the time domain. $J_c^{sq}$ and *f* are the amplitude and frequency of the charge current in the Pt film, respectively. *f* = 25 Hz in all LIT measurements. **c,** Image of the steady-state temperature *T* of the Pt/*Z*-cut $LiNbO_3$ sample with a Pt thickness $t_{Pt}$ of 12 nm and a $SiO_2$-spacer thickness of 7 nm. A coat of electrically insulating black ink enhances the infrared emissivity. White (yellow) dotted lines delimit the Pt ($SiO_2$) strip regions. **d,e,** Lock-in amplitude *A* (**d**) and phase $\phi$ (**e**) images of the Pt/*Z*-cut $LiNbO_3$ sample with $t_{Pt}$ = 12 nm at $J_c^{sq}$ = 8.0 mA with 512 and 340 pixels along the *x* and *y* directions, respectively. **f,g,** Line cuts of *A* (**f**) and $\phi$ (**g**) along the *x* direction of the images in **d** and **e**, respectively, obtained by averaging 60 raw profiles from the Pt area. L (R) is defined as the position at which the *A* signal exhibits the maximum value at the left (right) edge of the $SiO_2$ layer. **h,i,** $J_c^{sq}$ dependence of the *A* (**h**) and $\phi$ (**i**) signals at L (red) and R (blue). The error bars represent the standard deviation in the peak values of the raw profiles.

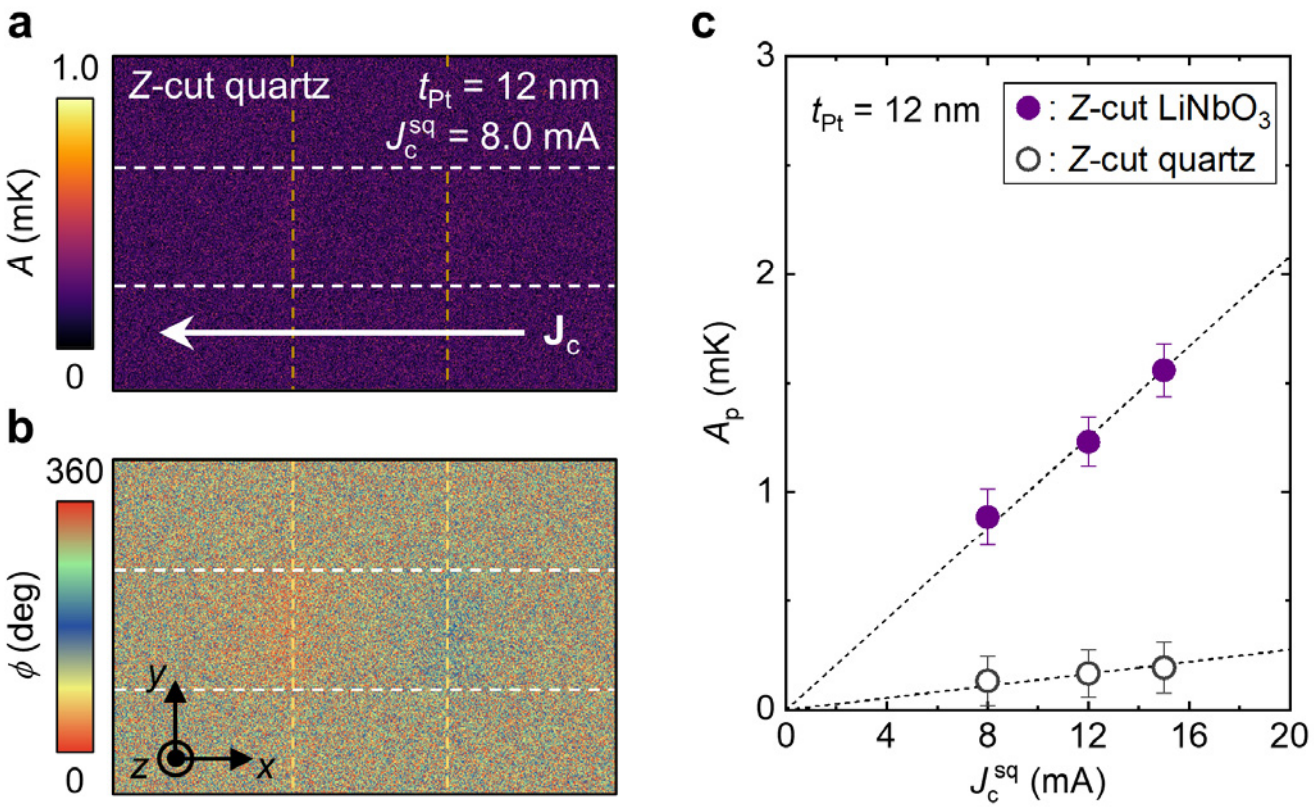

**Fig. 3 | Substrate dependence. a,b,** $A$ (**a**) and $\phi$ (**d**) images of the Pt/$Z$-cut quartz sample with $t_{\mathrm{Pt}}$ = 12 nm at $J_{\mathrm{c}}^{\mathrm{sq}}$ = 8.0 mA. **c,** $J_{\mathrm{c}}^{\mathrm{sq}}$ dependence of the peak amplitude $A_{\mathrm{p}}$, obtained by averaging the $A$ values at L and R, in the Pt/$Z$-cut $LiNbO_3$ sample (purple filled circles) and the Pt/$Z$-cut quartz sample (black open circles). The error bars represent the standard deviation in the $A_{\mathrm{p}}$ values of the 60 raw profiles along the $x$ direction between the white dotted lines in the thermal images.

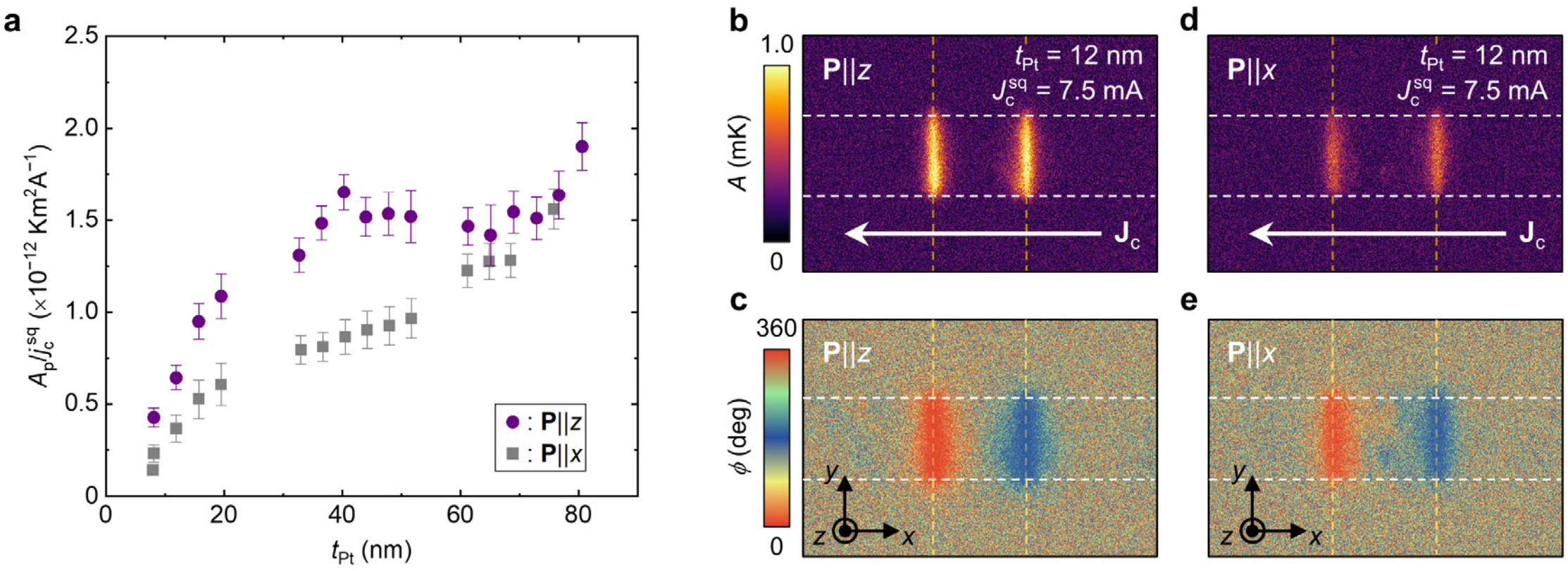

**Fig. 4 | Pt-thickness and polarization-direction dependences. a,** $t_{\mathrm{Pt}}$ dependence of $A_{\mathrm{p}}/j_{\mathrm{c}}^{\mathrm{sq}}$ for the Pt/$Z$-cut $LiNbO_3$ sample (purple filled circles) and Pt/$X$-cut $LiNbO_3$ sample (gray filled squares). $j_{\mathrm{c}}^{\mathrm{sq}}$ denotes the square-wave-modulated amplitude of the charge current density in the Pt film. The error bars represent the standard deviation in the $A_{\mathrm{p}}/j_{\mathrm{c}}^{\mathrm{sq}}$ values of the 60 raw profiles along the $x$ direction between the white dotted lines in the thermal images. **b,c,** $A$ (**b**) and $\phi$ (**c**) images for the Pt/$Z$-cut $LiNbO_3$ sample with $t_{\mathrm{Pt}}$ = 12 nm at $J_{\mathrm{c}}^{\mathrm{sq}}$ = 7.5 mA. **d,e,** $A$ (**d**) and $\phi$ (**e**) images for the Pt/$X$-cut $LiNbO_3$ sample with $t_{\mathrm{Pt}}$ = 12 nm at $J_{\mathrm{c}}^{\mathrm{sq}}$ = 7.5 mA.

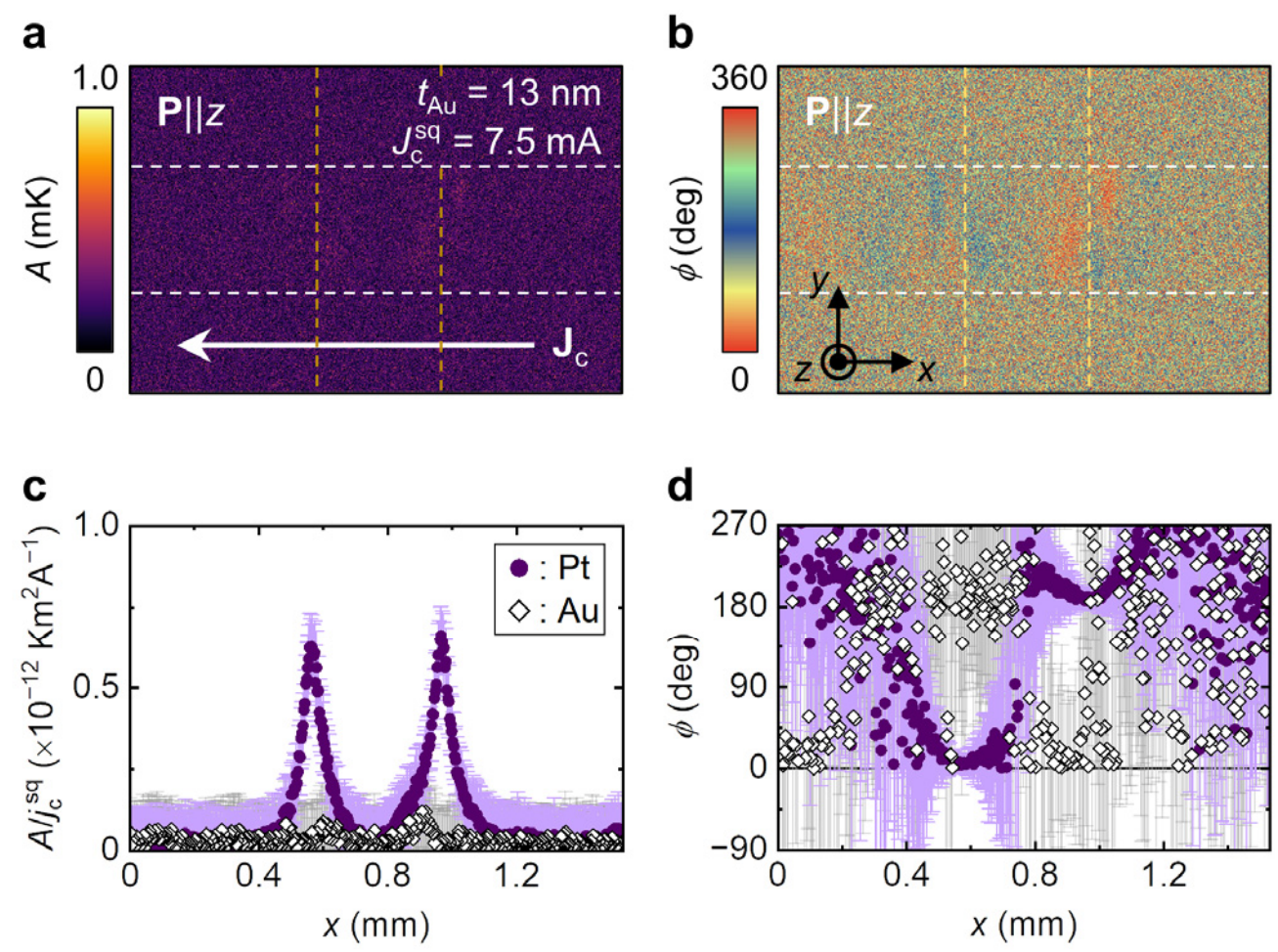

**Fig. 5 | Comparison between Pt/$LiNbO_3$ and Au/$LiNbO_3$. a,b,** $A$ (**a**) and $\phi$ (**b**) images for the Au/$Z$-cut $LiNbO_3$ sample with a Au thickness $t_{Au}$ of 13 nm at $J_c^{sq}$ = 7.5 mA. As in the Pt/$LiNbO_3$ samples, the $SiO_2$ strip is located between the vertical dashed orange lines. **c,d,** Line cuts of $A_p/j_c^{sq}$ (**c**) and $\phi$ (**d**) along the $x$ direction for the Pt/$Z$-cut $LiNbO_3$ sample (purple filled circles) and the Au/$Z$-cut $LiNbO_3$ sample (black open diamonds). The profiles for the Au/$Z$-cut $LiNbO_3$ sample were obtained from the images in **a** and **b**, respectively. The error bars represent the standard deviation in the $A_p/j_c^{sq}$ and $\phi$ values of the 60 raw profiles along the $x$ direction between the white dotted lines in the thermal images.

# Methods

## Sample preparation

The Pt/$LiNbO_3$ samples used for the experiments in Figs. 2 and 4 were prepared as follows. Before fabricating the Pt films, the $SiO_2$ films with a thickness gradient were grown on double-side polished single-crystal $LiNbO_3$ (0001) and (11-20) substrates (commercially available from Crystal Base Co., Ltd.), namely $Z$-cut and $X$-cut $LiNbO_3$ substrates, respectively. The $LiNbO_3$ substrates were ultrasonically cleaned in ethanol and acetone sequentially before deposition. The $SiO_2$ films were deposited from a stoichiometric $SiO_2$ target by radio-frequency magnetron sputtering using a combinatorial sputtering system (CMS-3200, Comet, Inc.) at room temperature. By moving a shutter near the substrate at a constant speed during film deposition, a linear thickness gradient was created. The process gas for the $SiO_2$ deposition consisted of Ar and $O_2$ with a volumetric flow ratio of 2:1 at a total working pressure of 0.4 Pa, where $O_2$ was introduced to enhance the insulating properties of the film. Metal masks with 6 slit openings with a width of 0.4 mm were put on the substrates during deposition, resulting in the fabrication of the $SiO_2$ strips with the same width and the long axis being perpendicular to the thickness gradient (Extended Data Fig. 1). Subsequently, the Pt films with a thickness gradient were deposited using the same sputtering system by direct-current magnetron sputtering in an Ar atmosphere at a pressure of 0.4 Pa at room temperature, employing the same metal masks rotated by 90° to form the Pt strips intersecting the $SiO_2$ strips after air exposure during mask replacement (Extended Data Fig. 1). $t_{\mathrm{Pt}}$ was determined by X-ray reflectivity using reference Pt films with the same thickness deposited without metal masks. The thickness gradients of the Pt and $SiO_2$ films were 3.3 nm/mm and 0.7 nm/mm, respectively. We confirmed that the small variation in film thickness across the width of the strips did not affect the LIT results. The Pt/$Z$-cut quartz sample used for the experiments in Fig. 3 and the Au/$LiNbO_3$ sample used for the experiments in Fig. 5 were prepared in the same manner as the Pt/$LiNbO_3$ samples. Before the LIT measurements, the top surface of the samples was coated with insulating black ink (JSC-3, Japansensor Corporation) having an emissivity of >0.94 to ensure high and uniform infrared emission.

For the resistivity measurements, the Pt films with the same thickness gradient were prepared either directly on $LiNbO_3$ or $SiO_2$/$LiNbO_3$. The Pt films in direct contact with $LiNbO_3$ were deposited without using metal masks and subsequently patterned into a Hall bar geometry by photolithography and Ar-ion milling, as shown in Extended Data Fig. 6a. For the Pt films on $SiO_2/LiNbO_3$, both $SiO_2$ and Pt films were deposited without using metal masks, followed by patterning the Pt layer into the Hall bar geometry using the same processes. Before depositing the Pt layer, the surface of the $SiO_2$ layer of the sample was exposed to air to maintain consistency with the samples used for LIT measurements. The longitudinal resistivity of the Pt films was measured using a four-probe method.

**Microstructure and composition analyses**

Scanning transmission electron microscopy (STEM) observations were conducted using a Titan G2 80-200 microscope (FEI Company) equipped with a probe aberration corrector and a Super-X energy-dispersive X-ray spectroscopy detector, operated at an accelerating voltage of 200 kV. Thin-foil specimens for STEM analysis were prepared using the lift-out method in a focused ion beam/scanning electron microscope (Helios G5, FEI Company).

**Thermal conductivity measurements**

The thermal conductivity $\kappa$ of the single-crystal $LiNbO_3$ and quartz substrates was measured by time-domain thermoreflectance (TDTR). All the substrates were coated with 70-nm-thick Al films using the sputtering system (CMS-3200, Comet, Inc.), because Al is a convenient transducer with a known thermoreflectance coefficient. Two mode-lock Er-doped fiber lasers were used as both the pump and probe laser beam sources with a repetition rate of 19.9952 MHz, pulse width of ~0.5 ps, and central wavelength for pump (probe) laser of 1550 nm (775 nm). The laser power and $1/e^2$ spot diameter were 16 mW (<1 mW) and 90 μm (30 μm) for the pump (probe) laser beam, respectively. The TDTR setup was configured with an electrical delay control system in which the oscillations of two lasers were synchronized using a function generator, ensuring time delay between pump and probe laser beams with 0-50 ns without a mechanical delay stage. The pump laser beam was modulated at the frequency of 0.2 MHz so that the temperature change at the sample surface was detected by the probe laser beam through a lock-in detection technique. To ensure reproducibility, the thermoreflectance was measured at three different spot positions on each sample, and the averaged signal was used for the analysis. Note that no spot position dependence was observed in all the samples. The lock-in phase of the thermoreflectance signal, corresponding to the ratio of in-phase and out-of-phase components of the signal, was fitted by a one-dimensional heat diffusion model and $\kappa$ of the substrates was estimated. The detailed experimental setup and analysis procedure of TDTR can be found elsewhere[46,47].

Extended Data Fig. 4a shows the TDTR signal for the *Z*-cut $LiNbO_3$, *X*-cut $LiNbO_3$, and *Z*-cut quartz substrates. As shown in the positive correlation of the sensitivity to $\kappa$ of the substrate (Extended Data Fig. 4b), which describes how strongly the TDTR signal responds to a change in a given parameter, the larger TDTR signal indicates higher $\kappa$.

**Numerical calculations**

Employing the method of finite-elements[48], we solved the heat diffusion equation for the temperature distributions generated by a pair of positive and negative heat sources, i.e., heat source and sink, induced by the thermoelectric effect. The temperature modulations were analyzed in the two-dimensional *x*-*z* cross-sections across the Pt strip (Extended Data Fig. 5a). The model consists of a Pt film with a thickness of 10 nm on a 0.4-mm-thick $LiNbO_3$ or 0.5-mm-thick quartz substrate and an overlying black ink layer with a thickness of 30 μm. The total width of the system along the *x* axis is 10 mm.

The heat diffusion equation was numerically solved in the frequency *f* domain: $2\pi i f C_v A e^{-i\phi} = \nabla(\kappa \nabla A e^{-i\phi}) + Q(r)$, where $C_v$ and $Q(r)$ are the volumetric specific heat and heat source, respectively. As schematically illustrated in Extended Data Fig. 5a, $Q(r)$ comprises a pair of the heat source $+Q$ and sink $-Q$ separated by an edge-to-edge distance of 0.4 mm. The heat source and sink having a width of 0.4 μm and thickness of 10 nm are arranged symmetrically in the substrate in the vicinity of Pt so that the midpoint between them is at the center of the model in the *x* direction. The system is thermally isolated except for the bottom of the substrate, which is assumed to be attached to a heat bath with a fixed temperature. The temperature modulation shown in Extended Data Fig. 5b is defined as the difference between the temperatures of the black ink surface directly above the centers of the heat source and sink. The temperature modulation at $f = 25$ Hz was calculated using $\kappa$ of the substrate as a parameter, where the $\kappa$ values of Pt and black ink are assumed to be 29.5 $Wm^{-1}K^{-1}$ (ref. 49) and 0.5 $Wm^{-1}K^{-1}$, respectively, and the $C_v$ values of Pt, $LiNbO_3$, quartz, and black ink are $2.79 \times 10^6$ $Jm^{-3}K^{-1}$ (ref. 50), 2.79

$\times 10^6$ $Jm^{-3}K^{-1}$ (ref. 51), $1.96 \times 10^6$ $Jm^{-3}K^{-1}$ (ref. 52), and $1.00 \times 10^6$ $Jm^{-3}K^{-1}$ (ref. 27), respectively.

**Theory of phonon-mediated nonlocal ferron-drag thermoelectricity**

Here we formulate the phonon-mediated nonlocal coupling between ferrons and electrons, as well as the associated ferron-drag thermoelectricity by perturbation theory for a bilayer composed of the ferroelectric $LiNbO_3$ and metallic Pt with the dielectric $SiO_2$ spacer. The direct Coulomb interaction between ferrons residing in the ferroelectric and electrons in the metal cannot explain the observations because it is efficiently screened, therefore weak, and short-ranged. An electron-ferron coupling mediated by acoustic phonons should suffer less from screening effects. Acoustic phonons are coherent throughout the entire bilayer system and therefore interact with both ferrons in the ferroelectric and electrons in the metal. In the ferroelectric layer, ferrons couple strongly to lattice distortions via the piezoelectric interaction of the form $\sim \varepsilon P$, where $\varepsilon$ denotes the strain field and $P$ the electric polarization. For convenience we formulate the problem in second quantization of the strain field and polarization fluctuations, noting that operators should be replaced by wave amplitudes in the classical regime. The ferron-phonon coupling takes the form:

$$H_{\mathrm{f-p}} = \sum_{q,\lambda} \eta_{q\lambda}(a_q + a_{-q}^{+})(b_{-q\lambda} + b_{q\lambda}^{+})$$

where $a_q^{+}(a_q)$ creates (annihilates) a ferron with a wave number $q$ in the ferroelectric and $b_{q\lambda}^{+}$ $(b_{-q\lambda})$ creates (annihilates) an acoustic phonon with $q$ and a branch (polarization) index $\lambda$. The coupling strength $\eta_{q\lambda}$ is in units of energy and scales with the ferroelectric volume $V_{\mathrm{F}}$ as $\eta_{q\lambda} \propto \sqrt{V_{\mathrm{F}}/V}$, where $V$ are the volume of the entire system, because the ferrons and piezoelectric coupling exist only in the ferroelectric layer. Electrons in the metallic layer couple to phonons via the conventional screened electron-phonon interaction:

$$H_{\mathrm{e-p}} = \sum_{q,\lambda} g_{kq\lambda} c_{k+q}^{+} c_k (b_{q\lambda} + b_{-q\lambda}^{+})$$

where $c_k^{+}$ $(c_k)$ is the creation (annihilation) operator of electrons with a wave number $k$. $g_{kq\lambda} \propto 1/\sqrt{V}$ denotes the electron-phonon coupling strength in the metallic layer, with the phonon amplitude normalized by $\sqrt{V}$. Since the phonon and ferron frequencies are very different, we may "integrate out" the conventional phonons by perturbation theory. We thus arrive at an effective phonon-mediated electron-

ferron interaction:

$$H_{\mathrm{e-f}} = \sum_{q,\lambda} V_{kq\lambda}^{\mathrm{e-f}} c_{k+q}^{+} c_k (a_q + a_{-q}^{\pm})$$

where $V_{kq\lambda}^{\mathrm{e-f}} = \sum_\lambda 2 g_{kq\lambda} \eta_{q\lambda} \omega_{q\lambda}^{\mathrm{p}} / [(\omega^2 - (\omega_{q\lambda}^{\mathrm{p}})^2]$ is the induced dynamic coupling strength with a phonon frequency $\omega_{q\lambda}^{\mathrm{p}}$. This coupling enables a net momentum transfer between an applied charge current and ferrons in the ferroelectric, thereby "dragging" along a heat-carrying ferron flow. The linearized Boltzmann equation in the relaxation time approximation[2] then leads to the ferron-drag coefficient $\Pi_{\mathrm{fd}}$, i.e., the ratio of the ferron-drag heat current in the ferroelectric to the applied charge current in the metal:

$$\Pi_{\mathrm{fd}} = \frac{2\pi e}{\sigma\hbar} \sum_{\boldsymbol{q}} \int \frac{d\boldsymbol{k}}{(2\pi)^3} |V_{kq\lambda}^{\mathrm{e-f}}|^2 \tau_q^{\mathrm{f}} \hbar\omega_q^{\mathrm{f}} (\hat{\boldsymbol{x}} \cdot \boldsymbol{v}_q^{\mathrm{f}}) \left[\tau_{k+q}^{\mathrm{e}} \left(\hat{\boldsymbol{x}} \cdot \boldsymbol{v}_{k+q}^{\mathrm{e}}\right) - \tau_k^{\mathrm{e}} (\hat{\boldsymbol{x}} \cdot \boldsymbol{v}_k^{\mathrm{e}})\right]$$

$$\frac{\partial n_q^{\mathrm{f}}}{\partial \hbar\omega_q^{\mathrm{f}}} \left(f_{k+q}^{\mathrm{e}} - f_k^{\mathrm{e}}\right) \delta\left(E_k - E_{k+q} + \hbar\omega_q^{\mathrm{f}}\right)$$

where $\boldsymbol{v}_q^{\mathrm{f}}$ ($\boldsymbol{v}_q^{\mathrm{e}}$) is the group velocity of ferrons (electrons) with energy $\hbar\omega_q^{\mathrm{f}}$ ($E_k$) and the equilibrium distribution $n_q^{\mathrm{f}}$ ($f_k^{\mathrm{e}}$) and $\tau_q^{\mathrm{f}} (\tau_k^{\mathrm{e}})$ is the momentum relaxation time of ferrons (electrons). $\sigma$ is the electrical conductivity of the metallic layer and $\hat{\boldsymbol{x}}$ is the direction of the applied charge current. Since $\sum_{\boldsymbol{q}} \to \int \frac{V}{(2\pi)^3} d\boldsymbol{q}$ and $V_{kq\lambda}^{\mathrm{e-f}} \propto \sqrt{V_{\mathrm{F}}/V}$, the ferron-drag coefficient is (inversely) proportional to the volume of the ferroelectric (the entire system): $\Pi_{\mathrm{fd}} \propto \Pi_0 V_{\mathrm{F}}/V$, where $\Pi_0$ is a constant ferron-drag coefficient independent of the Pt thickness. A dielectric spacer such as $SiO_2$ suppresses the ferron-drag contribution to the heat current by reducing the amplitude of the virtual phonons in the ferroelectric excited by the charge current and therefore the nonlocal electron-ferron coupling strength due to the amorphous-like structure of the spacer and Kapitza phonon scattering at the interfaces. Thus, the total thermoelectric coefficient in the metal/ferroelectric bilayer, which includes the conventional Peltier coefficient $\Pi_{\mathrm{e}}$, should be reduced when the metal is separated by the dielectric spacer, leading to heating and cooling at the intersection of these regions, consistent with our experiments using the Pt/$LiNbO_3$ samples (note that $\Pi_{\mathrm{e}}$ is unaffected by the ferroelectric). The temperature difference can be formulated by assuming that the $SiO_2$ spacer only increases the distance between the ferroelectric and metal without additional phonon scattering as

$$\Delta T = (\Pi_{\mathrm{fd}}^{\mathrm{Pt/LN}} - \Pi_{\mathrm{fd}}^{\mathrm{Pt/SO/LN}}) \frac{w_{\mathrm{SO}} t_{\mathrm{Pt}}}{K_{\mathrm{eff}}} j_{\mathrm{c}} \cong \Pi_0 \left\{\frac{t_{\mathrm{LN}}}{t_{\mathrm{LN}} + t_{\mathrm{Pt}}} - \frac{t_{\mathrm{LN}}}{t_{\mathrm{LN}} + t_{\mathrm{SO}} + t_{\mathrm{Pt}}}\right\} \frac{w_{\mathrm{SO}} t_{\mathrm{Pt}}}{K_{\mathrm{eff}}} j_{\mathrm{c}}$$

where $\Pi_{\rm fd}^{\rm Pt/LN}$ and $\Pi_{\rm fd}^{\rm Pt/SO/LN}$ represent the ferron-drag coefficient for the Pt/$LiNbO_3$ and Pt/$SiO_2$/$LiNbO_3$ regions, respectively, $w_{\rm SO}$ the width of the $SiO_2$ spacer, and $K_{\rm eff}$ the effective thermal conductance of the Pt/$SiO_2$/$LiNbO_3$ section. $t_{\rm SO}$ and $t_{\rm LN}$ denote the thicknesses of the $SiO_2$ spacer and $LiNbO_3$ substrate, respectively. In the second equality, we used a geometric scaling relation of $\Pi_{\rm fd}$. When the heat is conducted through the entire structure, $K_{\rm eff} = \kappa_{\rm SO} t_{\rm SO} + \kappa_{\rm Pt} t_{\rm Pt} + K_{\rm LN}$, where $K_{\rm LN} = \kappa_{\rm LN} t_{\rm LN}$ represents the thermal conduction contribution by the ferroelectric substrate. Since $t_{\rm LN}$ is much larger than $t_{\rm SO}$ and $t_{\rm Pt}$, $K_{\rm eff}$ is dominated by $K_{\rm LN}$ and $\Delta T$ scales approximately linearly with $t_{\rm Pt}$. This theory qualitatively reproduces the behavior in which the temperature changes due to the nonlocal ferron-drag effect increases monotonically with the Pt thickness, without affected by charge screening.

**Data availability**

The data that support the findings of this study are available from the corresponding author upon reasonable request.

**References**

[46] Yamazaki, T. et al. Quantitative measurement of figure of merit for transverse thermoelectric conversion in Fe/Pt metallic multilayers. *Phys. Rev. Appl.* **21**, 024039 (2024).

[47] Hirai, T. et al. Non-equilibrium magnon engineering enabling significant thermal transport modulation. *Adv. Funct. Mater.* **35**, 2506554 (2025).

[48] Hecht, F. New development in freefem++. *J. Numer. Math.* **20**, 251-265 (2012).

[49] Zhang, X. et al. Thermal and electrical conductivity of a suspended platinum nanofilm. *Appl. Phys. Lett.* **86**, 171912 (2005).

[50] Lu, L., Yi, W. & Zhang, D. L. 3ω method for specific heat and thermal conductivity measurements. *Rev. Sci. Instrum.* **72**, 2996-3003 (2001).

[51] Yao, S. et al. Growth, optical and thermal properties of near-stoichiometric $LiNbO_3$ single crystal. *J. Alloys. Compd.* **455**, 501-505 (2008).

[52] Grønvold, F., Stølen, S. & Svendsen, S. R. Heat capacity of α quartz from 298.15 to 847.3 K, and of β quartz from 847.3 to 1000 K—transition behaviour and revaluation of the thermodynamic. Thermochim. Acta **139**, 225-243 (1989).

**Acknowledgements** The authors thank M. Isomura, A. Kurita, E. Shimada, and S. Sekine for technical support. This work was supported by JSPS KAKENHI Grant-in-Aid for Scientific Research (S) (No. 22H04965, 24H02231), JST ERATO "Magnetic Thermal Management Materials" (No. JPMJER2201), and JST PRESTO "Information Carriers and Their Integrated Materials/Devices/Systems" (No. JPMJPR20B3). MANA is supported by World Premier International Research Center Initiative (WPI), MEXT, Japan.

**Author contributions** K.U. planned and supervised the study; T.I. and S.A. prepared the samples based on the design by R.I., T.H. and K.U.; T.I. and S.A. performed the lock-in thermography measurements; T.H. analyzed the thermography data, measured the thermal conductivity of the substrates, and prepared the figures; P.T., G.E.W.B. and K.U. developed the explanation of the experiments; H.S.A. performed the microstructure analysis of the samples; R.I. performed the numerical simulation; Y.K., T.S. and S.M. supported the sample preparation; K.U., P.T. and G.E.W.B. prepared the manuscript with input from T.I., T.H., H.S.A. and R.I. All the authors discussed the results and commented on the manuscript.

**Competing interests** The authors declare no competing interests.

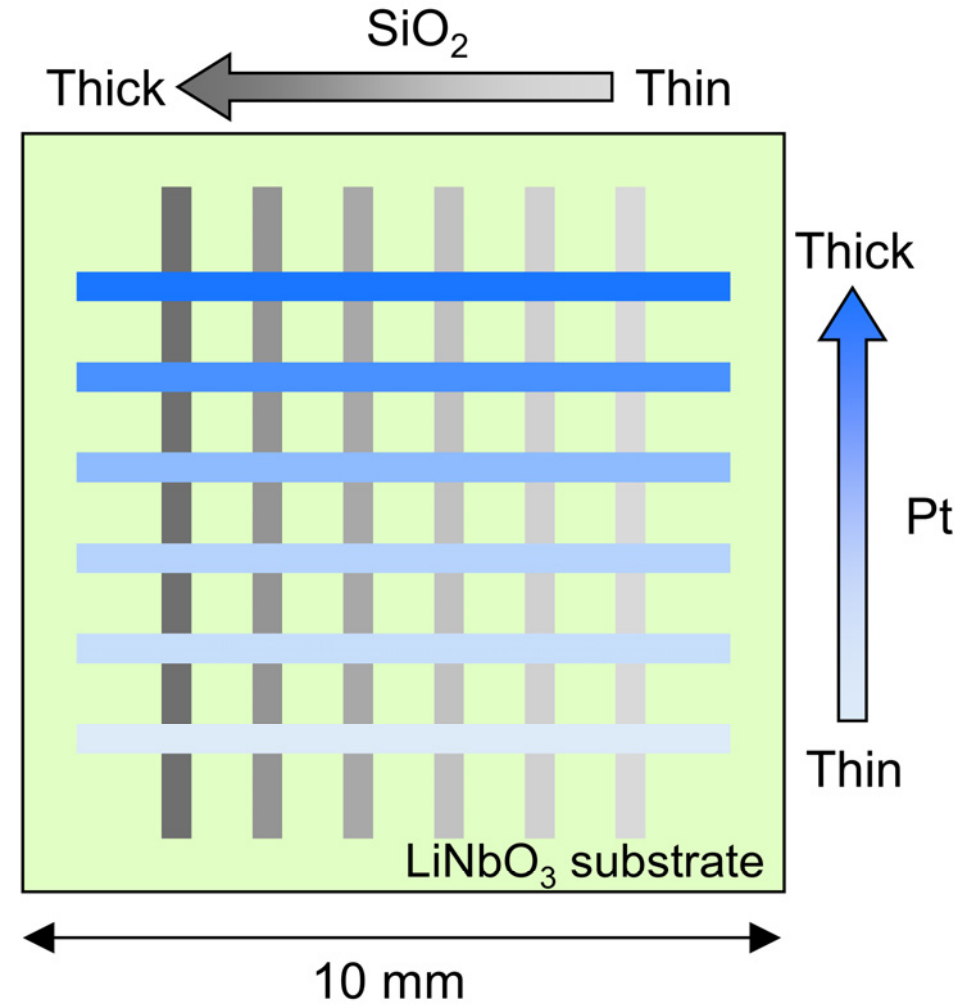

**Extended Data Fig. 1 | Sample structure.** Schematic of the Pt/$LiNbO_3$ sample with $SiO_2$ spacers used for the LIT measurements.

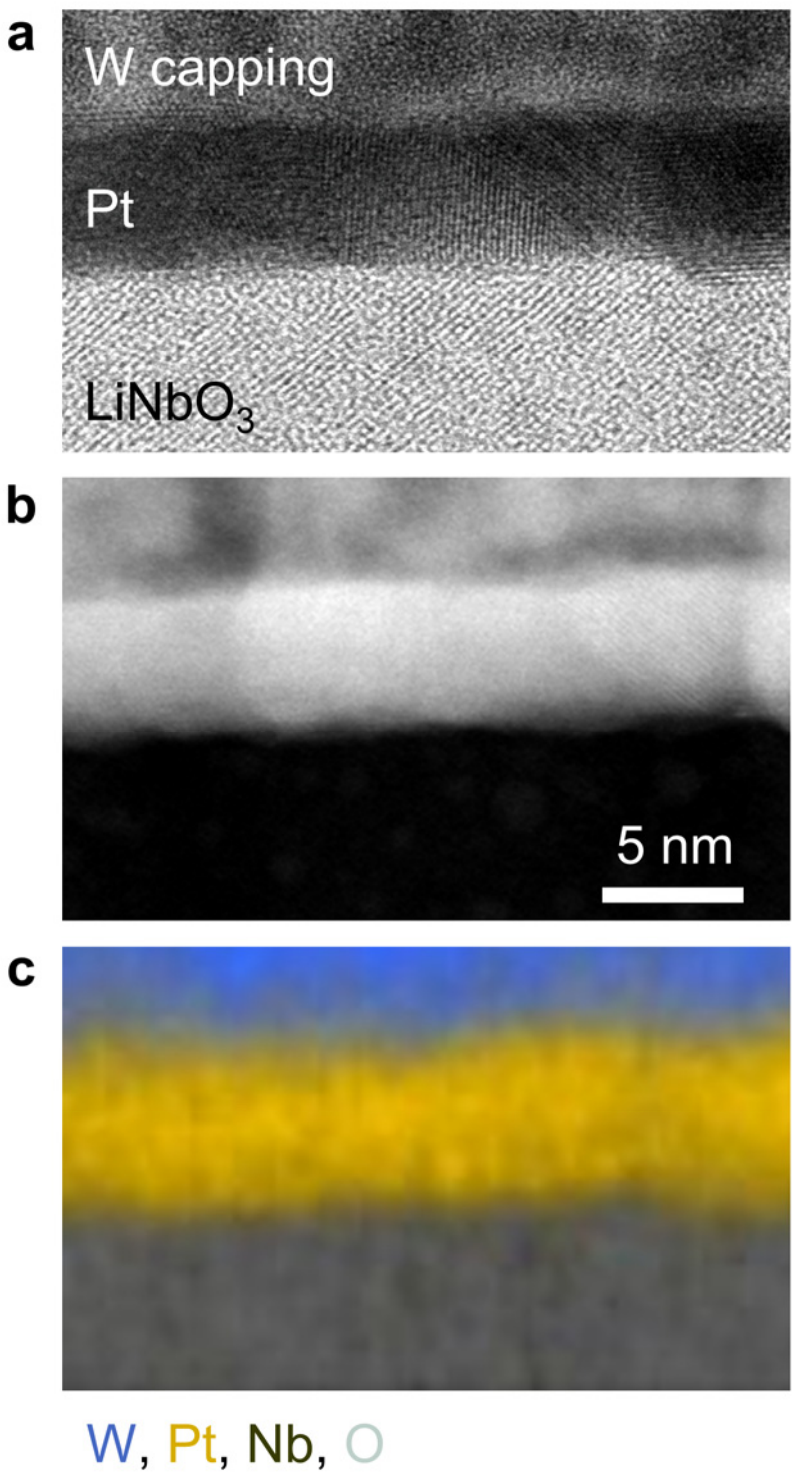

**Extended Data Fig. 2 | Interfacial microstructure of Pt/$LiNbO_3$. a,** Cross-sectional bright-field scanning transmission electron microscopy (STEM) image of the Pt/*Z*-cut $LiNbO_3$ stack, showing a continuous polycrystal Pt layer with a thickness of ~5.4 nm. A W capping was deposited as a protection layer. **b,** High-angle annular dark-field STEM image of the same region, confirming the uniform thickness and structural continuity of the Pt layer. **c,** Corresponding STEM energy-dispersive X-ray spectroscopy elemental map overlaid for W (blue), Pt (yellow), Nb (dark green), and O (light gray), revealing well-defined elemental distributions with minimal interdiffusion across the interfaces.

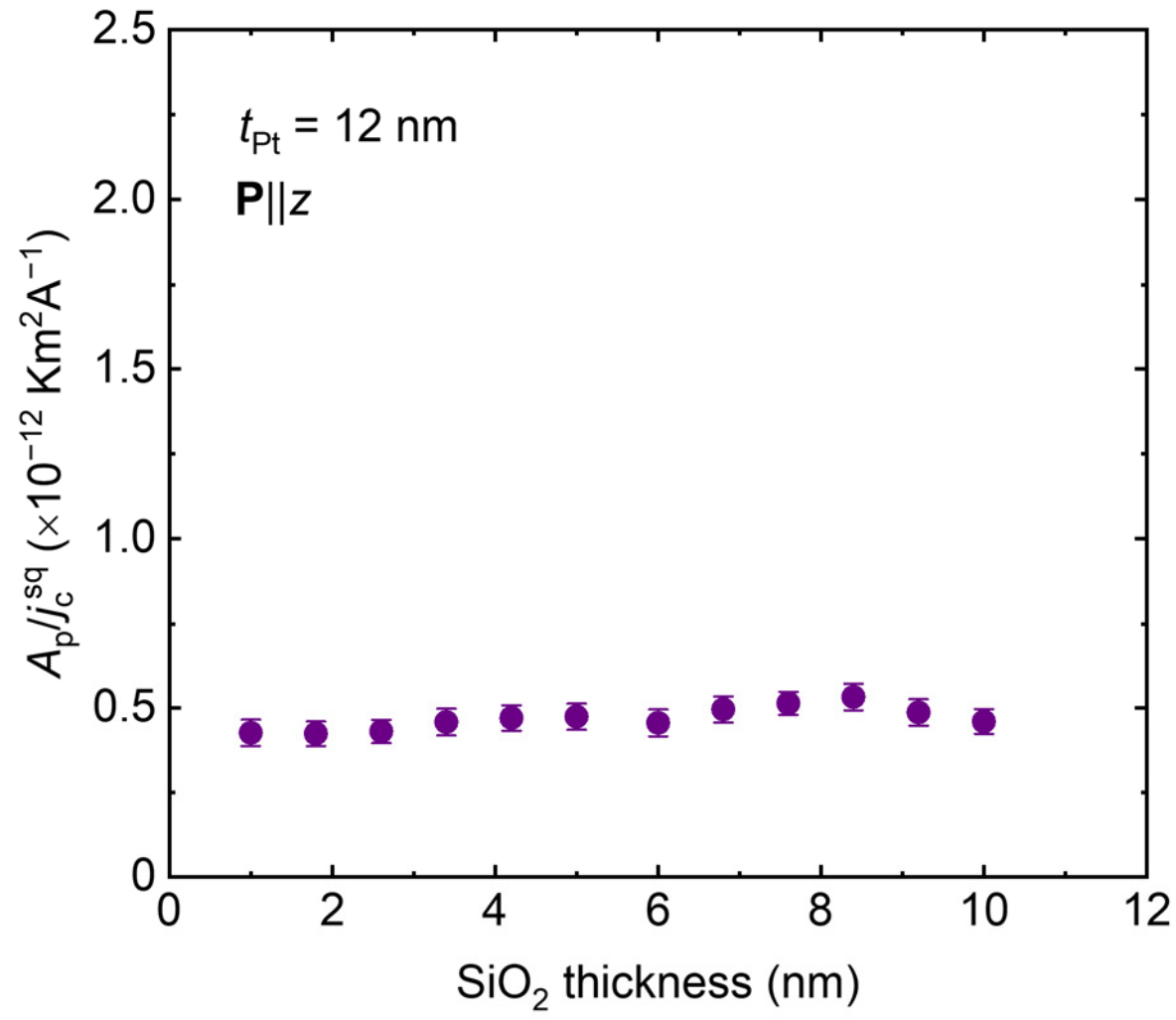

**Extended Data Fig. 3 | $SiO_2$-thickness dependence.** $A_{\mathrm{p}}/j_{\mathrm{c}}^{\mathrm{sq}}$ as a function of the thickness of the $SiO_2$ strip for the Pt/$Z$-cut $LiNbO_3$ samples. $t_{\mathrm{Pt}}$ was fixed at 12 nm in all samples. The error bars represent the standard deviation in the $A_{\mathrm{p}}/j_{\mathrm{c}}^{\mathrm{sq}}$ values of the 60 raw profiles along the $x$ direction in the thermal images.

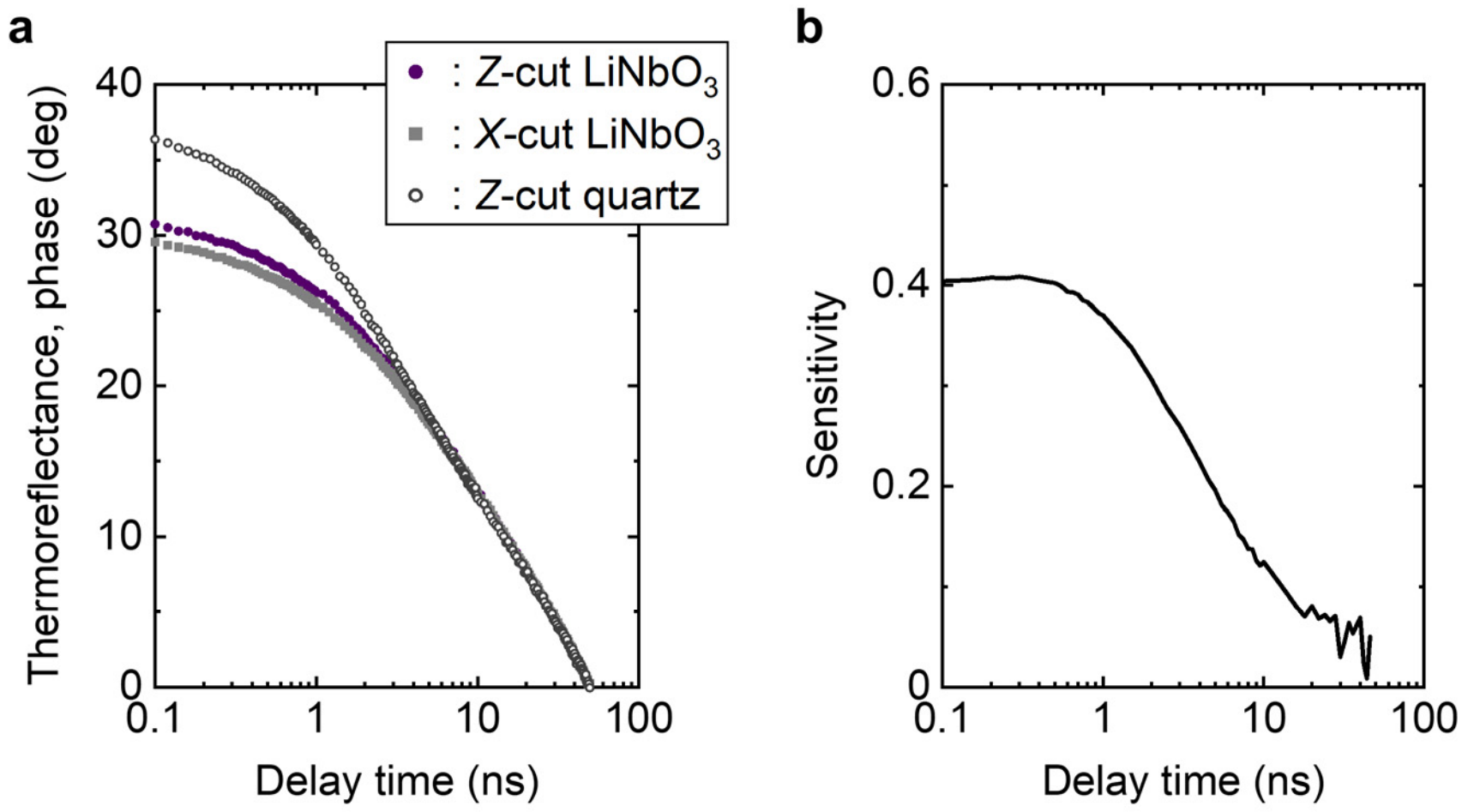

**Extended Data Fig. 4 | Time-domain thermoreflectance measurements. a,** Time-dependent thermoreflectance signals after pulsed laser excitation for the $Z$-cut $LiNbO_3$, $X$-cut $LiNbO_3$, and $Z$-cut quartz substrates coated with 70-nm-thick Al films. **b,** Sensitivity of the time-domain thermoreflectance signal to the thermal conductivity of the substrate. The sensitivity is defined as the logarithmic derivative of the time-domain thermoreflectance signal with respect to a fitting thermophysical parameter, and quantifies the relative change in the signal caused by a relative change in that parameter.

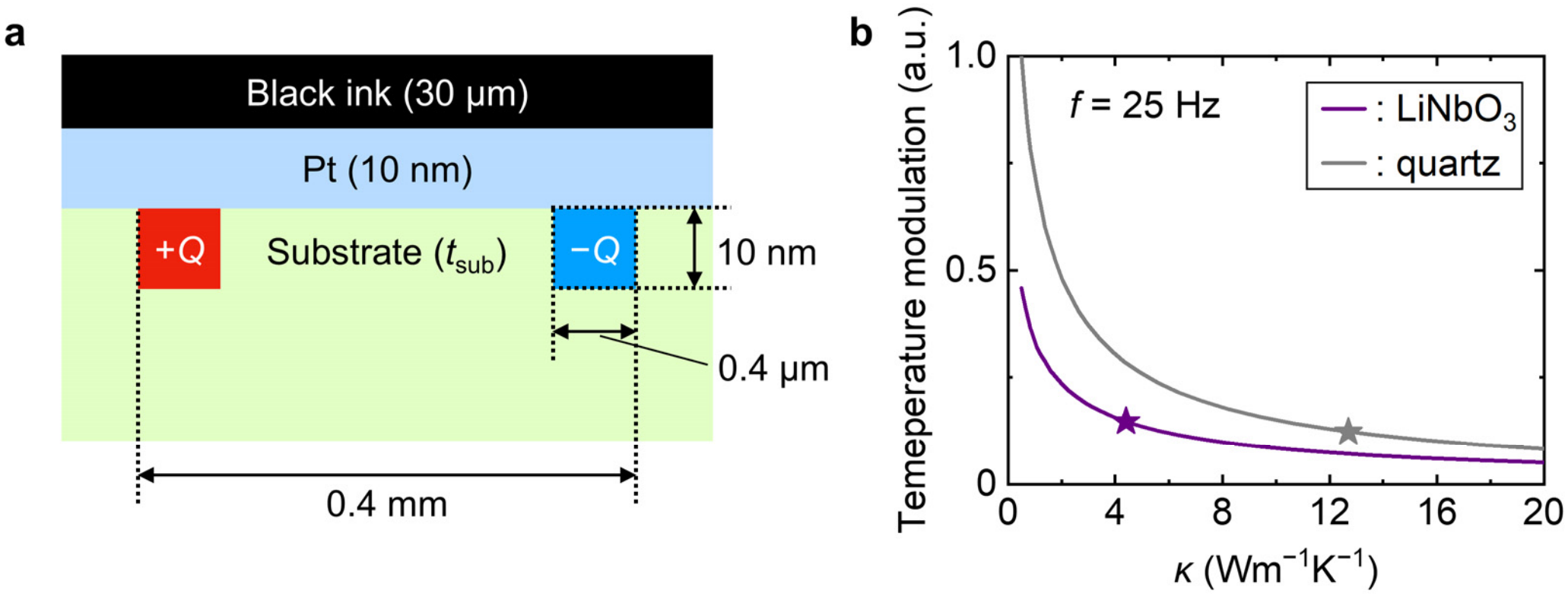

**Extended Data Fig. 5 | Effect of thermal conductivity of substrate on thermoelectric signals. a,** Schematic of the model geometry used for the finite-element calculations. $+Q$ ($-Q$) denotes the heat source (sink) induced in the substrate in the vicinity of the Pt layer. $t_{sub}$ denotes the thickness of the substrate. In accordance with the experimental condition, $t_{sub}$ was set to 0.4 mm (0.5 mm) for the $LiNbO_3$ (quartz) substrate. **b,** Dependence of the temperature modulation, i.e., temperature difference between the top surfaces above the heat source and sink, on the thermal conductivity $\kappa$ of the $LiNbO_3$ (purple curve) and quartz (gray curve) substrates at $f = 25$ Hz. The purple (gray) star data point shows the experimentally observed thermal conductivity value for the *Z*-cut $LiNbO_3$ (quartz) substrate; assuming the same $|Q|$, the magnitudes of the temperature modulation for both substrates are comparable because the effect of the difference in $\kappa$ was compensated by that in volumetric specific heat.

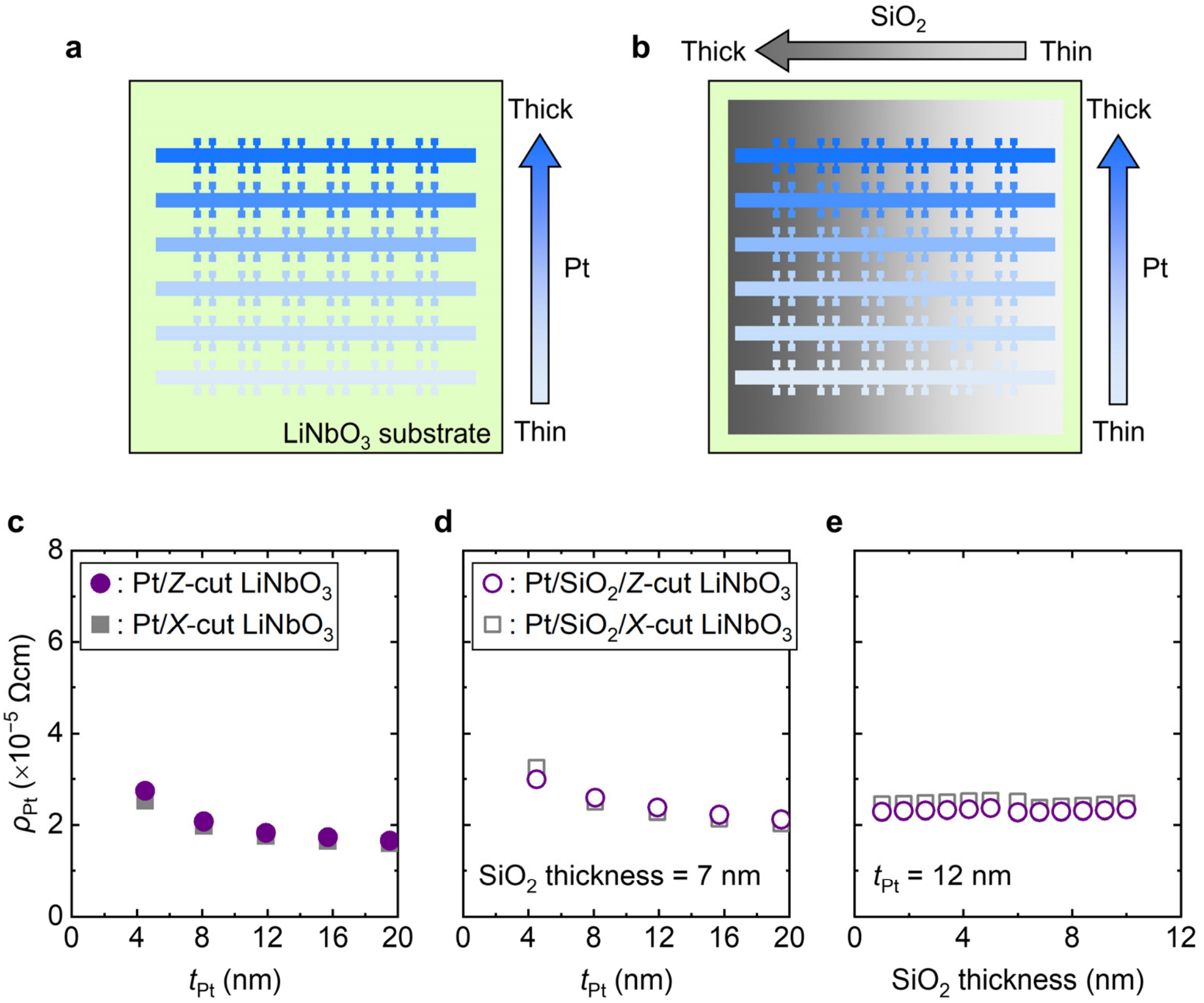

**Extended Data Fig. 6 | Electrical resistivity measurements. a,b,** Schematic of the Pt/$LiNbO_3$ samples without (**a**) and with (**b**) the $SiO_2$ layer used for the electrical resistivity measurements. Differently from the samples used for the LIT measurements, the Pt strips used here have small electrodes to measure the $t_{\mathrm{Pt}}$ and $SiO_2$-thickness dependences of the electrical resistivity $\rho_{\mathrm{Pt}}$ using the four-probe method. The Pt layer was patterned using photolithography and Ar-ion milling. Although the $SiO_2$ layer may also be removed by the milling process, this does not affect the $\rho_{\mathrm{Pt}}$ measurements. **c,d,** $t_{\mathrm{Pt}}$ dependence of $\rho_{\mathrm{Pt}}$ of the Pt films on the *Z*-cut $LiNbO_3$ substrates without (**c**) and with (**d**) the $SiO_2$ interlayer. In **d,** the $SiO_2$ thickness was fixed at 7 nm. **e,** $SiO_2$-thckness dependence of $\rho_{\mathrm{Pt}}$ of the 12-nm-thick Pt film on the $SiO_2$-coated *Z*-cut $LiNbO_3$ substrate.